\documentclass[aps,prb,twocolumn,superscriptaddress,floatfix]{revtex4-1}

\usepackage{bm}
\usepackage{amsmath,amssymb,amsfonts}
\usepackage{graphicx}
\usepackage{natbib}
\usepackage{dcolumn}
\usepackage{bbold}

\newcommand{\abs}[1]{\lvert #1 \rvert}

\newcommand{\mean}[1]{\langle #1 \rangle}
\newcommand{\ket}[1]{\lvert #1 \rangle}
\newcommand{\bra}[1]{\langle #1 \rvert}
\newcommand{\expval}[3]{\bra{#1}\hat{#2}\ket{#3}}
\newcommand{\expvalnh}[3]{\bra{#1}#2\ket{#3}}
\newcommand{\overlap}[2]{\langle #1 \lvert #2 \rangle}
\newcommand{\operator}[3]{\ket{#1} #2 \bra{#3}}
\newcommand{\idop}{\hat{\mathbb{1}}}

\newcommand{\mct}[1]{\multicolumn{1}{c}{#1}}

\begin{document}
\title{The Quantum Energy Density: Improved Efficiency for Quantum Monte Carlo}

\author{Jaron~T.~Krogel}
\affiliation{Department of Physics, University of
Illinois at Urbana-Champaign, Urbana, IL 61801, USA}

\author{Min~Yu}
\affiliation{Molecular Foundry, Lawrence Berkeley National Laboratory, Berkeley, CA 94720, USA}

\author{Jeongnim~Kim}
\affiliation{Materials Science and Technology Division and Computational Chemistry and Materials Division, Oak Ridge National Laboratory, Oak Ridge, TN 37831, USA}

\author{David~M.~Ceperley}
\affiliation{Department of Physics, University of
Illinois at Urbana-Champaign, Urbana, IL 61801, USA}

\begin{abstract}
We establish a physically meaningful representation of a quantum energy density for use in Quantum Monte Carlo calculations. The energy density operator, defined in terms of Hamiltonian components and density operators, returns the correct Hamiltonian when integrated over a volume containing a cluster of particles. This property is demonstrated for a helium-neon ``gas,'' showing that atomic energies obtained from the energy density correspond to eigenvalues of isolated systems. The formation energies of defects or interfaces are typically calculated as total energy differences. Using a model of delta-doped silicon (where dopant atoms form a thin plane) we show how interfacial energies can be calculated more efficiently with the energy density, since the region of interest is small. We also demonstrate how the energy density correctly transitions to the bulk limit away from the interface where the correct energy is obtainable from a separate total energy calculation.
\end{abstract}

\maketitle

\section{Introduction}\label{intro}

A description of local energetics at the quantum level opens the possibility of greater insight into properties as fundamental as bonding or energy transport in matter.  In electromagnetism, formulations of the energy density are well known, but quantum analogues of the classical representations have received only limited attention for many-body condensed matter systems.  As in the classical setting, representations of the quantum energy density are not unique, but this ambiguity is mitigated somewhat by the fact that each representation is an observable of the quantum system.

The search for a more complete understanding of energy transport in phonon systems led Hardy\cite{har63} to define a quantum energy density operator, closely related to the Hamiltonian, nearly 50 years ago.  The perspective taken in that work was to associate distinct energies with each quantum particle.  Later on, and perhaps independently, Ziesche and Lehmann\cite{zie87} introduced alternative representations of the kinetic energy density as well as a field form of the potential energy for both the many-body wavefunction and Kohn-Sham\cite{kohn1965} orbitals.  Similar forms were also recognized by Chetty and Martin,\cite{che92} and have recently been applied\cite{yu11} to the problem of defect formation energies and surface relaxations, in the context of Density Functional Theory\cite{hohenberg1964} (DFT).  

In this work, we expand on prior formulations of the quantum energy density and investigate its application within Quantum Monte Carlo (QMC) calculations, which allow access to the many-body properties of quantum systems with high accuracy.  Following the work of Hardy, we seek quantum operators to represent observable forms of the energy density.  A new form of the kinetic energy density operator is presented, disrupting the notion of unique kinetic energies within specialized atomic volumes.\cite{yu11}  If the perspective is taken that particles carry the energy, we also show that the partitioning of potential energy among various charge species is not arbitrary, but must take a unique form if long-ranged energy transfers between neutral subsystems are to be avoided.  Appropriate forms of the energy density operator for use in Diffusion Monte Carlo\cite{grimm1971,anderson1976} (DMC) simulations are established with special consideration given to typical use cases such as twist averaging.\cite{lin2001}  As the energy operator for a subsystem does not, in general, commute with the Hamiltonian, we compare the appropriateness of mixed versus extrapolated estimates of subsystem energies.

A great advantage of such a local description of energetics is that it offers an alternative to the ``brute force'' approach of calculating formation energies from total energy differences, which is currently the standard way of obtaining formation energies of surfaces, interfaces, and defects. A large portion of this computational effort is extraneous due to cancellation between similar parts of the systems.  For Quantum Monte Carlo, the situation is even worse because these similar regions contribute substantially to the statistical variance while adding nothing to the final answer.  The use of densities reduces this problem, since quantities are accumulated over the area of interest alone, which eliminates the noise from other regions.  The efficiency gains of this approach can be quite substantial, especially if the region of interest is small compared to the total simulation volume, a property we demonstrate with a model of $\delta$-doped silicon.

The remainder of the paper is organized as follows.  In section \ref{sec_derivation} we derive kinetic and potential energy densities (along with corresponding operators) starting from standard expressions for the respective energies.  A single form of the energy density operator is recommended for use in DMC.
We then validate our chosen form of the energy density operator in section \ref{sec_HeNe} with a model of a helium-neon gas.  This system is chosen because the atoms involved are weakly interacting, providing separable atomic energies.  Efficiency enhancements for QMC energy difference calculations are explored in detail in section \ref{sec_ddSi} with a simple model of $\delta$-doped silicon, where dopant atoms are confined to a single crystal plane.  In section \ref{sec_dft}, we find the Kohn-Sham energy density resulting from our representation and demonstrate its close relationship with one recently employed in DFT by Yu, Trinkle, and Martin.\cite{yu11}  We also compare atomic energies of $\delta$-doped silicon obtained from the Martin form in DFT with Diffusion Monte Carlo results.  Our major conclusions are then summarized in section \ref{sec_conclusion}.  Appendix \ref{altderiv} contains an alternative derivation of the energy density operator starting from a general form and imposing successive physical conditions.  Potential energy densities in Ewald\cite{ewald1921} systems are treated in appendix \ref{ewalddens}.

\section{Derivation of the Energy Density}\label{sec_derivation}
All descriptions of energy density ($\mathcal{E}_r$) must satisfy the simple requirement
\begin{align}\label{normalization}
  E = \int dr \mathcal{E}_r,
\end{align} 
where $E$ is the total energy.  While Eq. \ref{normalization} is only loosely constraining, it can be used to arrive at fairly natural forms for the energy density.  When these forms are recast in terms of energy density operators, it becomes clear that this ``natural'' approach amounts to preserving the exchange symmetries of the Hamiltonian in each corresponding energy density operator.  Even within such highly symmetric forms, ambiguity remains.  Valid forms of the energy density are evaluated on the basis of suitability for use in QMC calculations.

\subsection{Kinetic Energy Density}
The hermiticity of the momentum operator immediately provides three equivalent forms for the kinetic energy,
\begin{align}
  T &= -\frac{1}{2}\sum_i\overlap{\nabla_i^2\Psi}{\Psi} \\
    &=  \frac{1}{2}\sum_i\overlap{\nabla_i\Psi}{\nabla_i\Psi} \\
    &= -\frac{1}{2}\sum_i\overlap{\Psi}{\nabla_i^2\Psi} ,
\end{align}
where the sum is over all quantum particles.  Using exchange symmetries and transforming to real space, these forms become
\begin{align}
  T &= -\frac{N}{2}\int dR \nabla_1^2\Psi(R)^*\Psi(R) \\
    &=  \frac{N}{2}\int dR \nabla_1\Psi(R)^*\nabla_1\Psi(R)\\
    &= -\frac{N}{2}\int dR \Psi(R)^*\nabla_1^2\Psi(R),
\end{align}
where $R$ represents the $N$ particle coordinate vector, $R=[r_1,r_2,\ldots,r_N]$.  For notational convenience, we introduce $R_{\setminus_i}$ which is just $R$ excluding $r_i$.

In the spirit of the number density, $\rho_r\equiv N\int dR_{\setminus_1}\abs{\Psi(r,R_{\setminus_1})}^2$, we recognize distinct kinetic energy densities
\begin{align}
  T^{L^2}_r &= -\frac{N}{2}\int dR_{\setminus_1} \nabla_r^2\Psi(r,R_{\setminus_1})^*\Psi(r,R_{\setminus_1}) \\
  T^{LR}_r  &=  \frac{N}{2}\int dR_{\setminus_1} \nabla_r\Psi(r,R_{\setminus_1})^*\nabla_r\Psi(r,R_{\setminus_1})\\
  T^{R^2}_r &= -\frac{N}{2}\int dR_{\setminus_1} \Psi(r,R_{\setminus_1})^*\nabla_r^2\Psi(r,R_{\setminus_1}).
\end{align}
The superscripts denote the direction of action of the gradients in the expectation value, left ($L$) or right ($R$).  $T^{L^2}_r$ and $T^{R^2}_r$ have the undesirable feature of being complex valued.  After separating the real and imaginary parts, the relationship between the three energy densities takes a more illuminating form with help from the chain rule,
\begin{align}
  T^{R^2}_r-T^{L^2}_r &= \frac{i}{2}\nabla_r\cdot j_r \\
  \frac{T^{R^2}_r+T^{L^2}_r}{2} &= T^{LR}_r-\frac{1}{4}\nabla_r^2\rho_r,
\end{align}
where $\rho_r$ and $j_r$ are the number and current densities, respectively.

To construct operators corresponding to the above densities, we insert density operators within the standard operator for the total kinetic energy,
\begin{align}
  \hat{T} = \frac{1}{2}\sum_i\hat{p}_i^2.
\end{align}
The real space density operator for particle $i$ is defined as
\begin{align}
  \hat{\delta}_{rr_i}=\int dR\operator{R}{\delta_{rr_i}}{R}.
\end{align}
Since each density operator integrates to the identity,
\begin{align}
  \int dr \hat{\delta}_{rr_i} = \int dR\operator{R}{}{R}\equiv \idop,
\end{align}
any insertion of density operators into $\hat{T}$ will create a kinetic energy density operator $\hat{T}_r^X$ satisfying the normalization condition
\begin{align}
  \int dr \hat{T}_r^X = \hat{T}.
\end{align}
The operators corresponding to $T^{L^2}_r$, $T^{LR}_r$, and $T^{R^2}_r$ are then given by
\begin{align}
  \hat{T}^{L^2}_r &= \frac{1}{2}\sum_i\hat{p}_i^2\hat{\delta}_{rr_i} \\ 
  \hat{T}^{LR}_r  &= \frac{1}{2}\sum_i\hat{p}_i\hat{\delta}_{rr_i}\hat{p}_i \\ 
  \hat{T}^{R^2}_r &= \frac{1}{2}\sum_i\hat{\delta}_{rr_i}\hat{p}_i^2,
\end{align}
as can be confirmed directly by calculating $\expvalnh{\Psi}{\hat{T}_r^X}{\Psi}$.  The complex nature of $T^{L^2}_r$ and $T^{R^2}_r$ surfaces here in the non-hermiticity of the corresponding operators.  Again, a symmetric combination remedies the problem:
\begin{align}
  \hat{T}^{L^2+R^2}_r &= \frac{1}{2}\sum_i\frac{\hat{p}_i^2\hat{\delta}_{rr_i}+\hat{\delta}_{rr_i}\hat{p}_i^2}{2}.
\end{align}

Density operator insertion also reveals a less obvious ``modulus'' form:
\begin{align}
  \hat{T}^{mod}_r  &= \frac{1}{2}\sum_i\abs{\hat{p}_i}\hat{\delta}_{rr_i}\abs{\hat{p}_i} .
\end{align}
Though difficult to apply in real space, this form is perfectly reasonable in a momentum space formulation since $\abs{\hat{p}}\ket{p}=\abs{p}\ket{p}$.  Since $T^{mod}_r$ cannot be related to the other forms of the kinetic energy densities by introducing a vanishing surface term, integration over ``gauge independent''\cite{yu11} volumes will not provide unique kinetic energies.

In Diffusion Monte Carlo calculations, expectation values of observables are approximated by ``mixed'' estimates:
\begin{align}\label{mixed_estimator}
  \mean{A}\approx \frac{\expval{\Psi_0}{A}{\Psi_T}}{\overlap{\Psi_0}{\Psi_T}}.
\end{align}
Here $\hat{A}$ is the observable in question, $\Psi_T$ is a variational approximation of the ground state, and $\Psi_0$ is the exact ground state (for fermions it is actually the lowest energy state sharing the nodes\cite{anderson1975,anderson1976} or phase\cite{ortiz1993} of $\Psi_T$).  Any leftward ($L$) acting representations of the kinetic energy operator are challenging for DMC, since they involve derivatives of $\Psi_0$, and hence derivatives of the projection operator used to obtain $\Psi_0$ from $\Psi_T$.  It would therefore be ideal if $\hat{T}^{R^2}_r$ could be used to obtain the energy density.  Fortunately, there is a class of situations that make this choice possible.

In order to reduce finite size effects, observables in QMC calculations of extended systems are often calculated with twist averaged boundary conditions (TABC), \emph{i.e.} they are integrated over the first Brillouin zone of the simulation cell.  At each $k$-point, solutions ($\Psi_k$) to the Schrodinger equation\cite{schrodinger1926} satisfy a many-body form\cite{rajagopal1994,rajagopal1995} of the standard\cite{bloch1928} Bloch condition, and hence $\Psi_{-k}=\Psi_k^*$.  From this, we can relate $T^{L^2}_{r,k}$ and  $T^{R^2}_{r,k}$:
\begin{align}
  T^{R^2}_{r,-k} &= \expvalnh{\Psi_{-k}}{\hat{T}^{R^2}_r}{\Psi_{-k}} \nonumber\\
               &= \expvalnh{\Psi_{k}}{\hat{T}^{R^2\dagger}_r}{\Psi_{k}} \nonumber\\
               &= \expvalnh{\Psi_{k}}{\hat{T}^{L^2}_r}{\Psi_{k}} \nonumber\\
               &= T^{L^2}_{r,k}.
\end{align}
If the $k$-point set used to approximate the Brillouin zone integration has inversion symmetry, we further see that
\begin{align}
  T^{R^2}_{r,TABC} &= \frac{1}{N_k}\sum_k T^{R^2}_{r,k} \nonumber\\
                 &= \frac{1}{N_k}\sum_k \frac{T^{R^2}_{r,k}+T^{R^2}_{r,-k}}{2} \nonumber\\
                 &= \frac{1}{N_k}\sum_k \frac{T^{R^2}_{r,k}+T^{L^2}_{r,k}}{2} \nonumber\\
                 &= T^{L^2+R^2}_{r,TABC}.
\end{align} 
Thus $T^{R^2}_{r,TABC}$ will be real-valued for pure (non-mixed) estimators as long as inversion symmetry of the $k$-point grid is maintained.  For DMC, the mixed estimator (see Eq. \ref{mixed_estimator}) will again yield a complex-valued kinetic energy density, but since the imaginary part is due solely to errors in the trial function $\Psi_T$, it can safely be ignored.  In the remainder of the paper, $\hat{T}^{R^2}_r$ is used as the kinetic energy density operator.

\subsection{Potential Energy Density}
When deriving the potential energy density, we consider a system comprised of electrons (neglecting spin) and identical ions with nuclear charge $Z$.  This choice is made to simplify the analysis while maintaining the essential partitioning between fast moving particles (the electrons) and slow moving or immobile particles (the ions).  Generalizing to multiple ionic species, or any other system interacting via pair potentials, is straightforward.  The potential energy operator of such a system is given by
\begin{align}\label{totalpot}
  \hat{V} = \frac{1}{2}\sum_{i\ne j}\hat{v}^{ee}_{r_ir_j} + \sum_{i\ell}\hat{v}^{eI}_{r_i\tilde{r}_\ell} + \frac{1}{2}\sum_{\ell\ne m}\hat{v}^{II}_{\tilde{r}_\ell\tilde{r}_m},
\end{align}
where $v^{ee}$, $v^{eI}$, and $v^{II}$ are the electron-electron, electron-ion, and ion-ion Coulomb energies, respectively.  In the above and in following expressions, electron coordinates are denoted $r_i$ or $r_j$ while ion coordinates are denoted $\tilde{r}_\ell$ or $\tilde{r}_m$.

In order to translate the potential energy into a density, we enlist the help of the pair density operator and its factorization according to electron or ion species,
\begin{align}
  \hat{\rho}_{rr'} &= \sum_{p\ne p'}\hat{\delta}_{rr_p}\hat{\delta}_{r'r_p'}\\
                  &= \sum_{i\ne j}\hat{\delta}_{rr_i}\hat{\delta}_{r'r_j} 
                    +\sum_{i\ell}\hat{\delta}_{rr_i}\hat{\delta}_{\tilde{r}'\tilde{r}_\ell} \nonumber\\
                  &\quad  +\sum_{i\ell}\hat{\delta}_{r\tilde{r}_\ell}\hat{\delta}_{r'r_i}
                    +\sum_{\ell\ne m}\hat{\delta}_{r\tilde{r}_\ell}\hat{\delta}_{r'\tilde{r}_i}\\
                  &= \hat{\rho}^{ee}_{rr'} + \hat{\rho}^{e}_{r}\hat{\rho}^I_{r'} + \hat{\rho}^{I}_{r}\hat{\rho}^e_{r'} + \hat{\rho}^{II}_{rr'}, 
\end{align}
where $\hat{\rho}^e_r$ and $\hat{\rho}^I_r$ are the single particle density operators
\begin{align}
  \hat{\rho}^e_r = \sum_i \hat{\delta}_{rr_i} \qquad
  \hat{\rho}^I_r = \sum_\ell \hat{\delta}_{r\tilde{r}_\ell}.
\end{align} 
By rewriting the pair potential operators in the form
\begin{align}
  \hat{v}_{r_ir_j} = \int drdr' \hat{\delta}_{rr_i}\hat{\delta}_{rr_j}\hat{v}_{rr'},
\end{align}
the total potential energy becomes
\begin{align}
  V &= \expval{\Psi}{V}{\Psi} \\
    &= \frac{1}{2}\int drdr'\Big[\rho^{ee}_{rr'}v^{ee}_{rr'}+\rho^{e}_{r}\rho^I_{r'}v^{eI}_{rr'} \nonumber\\
    & \qquad\qquad\qquad+\rho^{I}_{r}\rho^e_{r'}v^{Ie}_{rr'}+\rho^{II}_{rr'}v^{II}_{rr'}\Big].
\end{align}
In this step, we have made use of the Born-Oppenheimer approximation,\cite{born1927} so that $\rho^{eI}_{rr'}\approx\rho^e_r\rho^I_{r'}$ with $\rho^e_r$ depending on ion positions.
This immediately suggests potential energy densities ``belonging'' to electrons or ions:
\begin{align}
  V^e_r &= \frac{1}{2}\int dr'\left[\rho^{ee}_{rr'}v^{ee}_{rr'}+\rho^{e}_{r}\rho^I_{r'}v^{eI}_{rr'}\right] \\
  V^I_r &= \frac{1}{2}\int dr'\left[\rho^{I}_{r}\rho^e_{r'}v^{Ie}_{rr'}+\rho^{II}_{rr'}v^{II}_{rr'}\right].
\end{align}
The separation of terms is made on the basis of whether $r$ is equal (via a delta function) to an electron or ion coordinate.

In many electronic structure calculations, including the majority of QMC calculations, the ions are considered to be classical and immobile.  In this case, densities for the ion terms revert to collections of delta functions,
\begin{align}
  \rho^I_r = \sum_\ell\delta_{r\tilde{r}_\ell} \qquad \rho^{II}_{rr'}=\sum_{\ell\ne m}\delta_{r\tilde{r}_\ell}\delta_{r'\tilde{r}_m},
\end{align}
and so the electron and ion potential energy densities become
\begin{align}
  V^e_r &= \frac{1}{2}\int dr' \rho^{ee}_{rr'}v^{ee}_{rr'}+\frac{1}{2}\rho^e_r\sum_\ell v^{eI}_{r\tilde{r}_\ell} \label{Ver}\\
  V^I_r &= \frac{1}{2}\sum_\ell\delta_{r\tilde{r}_\ell}\int dr'\rho^e_{r'}v^{Ie}_{rr'}+\frac{1}{2}\sum_\ell\delta_{r\tilde{r}_\ell}\sum_{m\ne\ell}v^{II}_{\tilde{r}_\ell\tilde{r}_m}\label{Vir}.
\end{align}

When rewritten as an operator, the potential energy density takes on the simple form
\begin{align}\label{pdensop_part}
  \hat{V}_r &= \sum_{i<j}\frac{\hat{\delta}_{rr_i}+\hat{\delta}_{rr_j}}{2}\hat{v}^{ee}_{r_ir_j}
              + \sum_{i\ell}\frac{\hat{\delta}_{rr_i}+\hat{\delta}_{r\tilde{r}_\ell}}{2}\hat{v}^{eI}_{r_i\tilde{r}_\ell}\nonumber\\ 
   & \qquad   + \sum_{\ell< m}\frac{\hat{\delta}_{r\tilde{r}_\ell}+\hat{\delta}_{r\tilde{r}_m}}{2}\hat{v}^{II}_{\tilde{r}_\ell\tilde{r}_m}.
\end{align}
Thus each particle carries half the energy of any pair interaction in which it participates.  The same partitioning was also chosen by Hardy, but it was presented as a somewhat arbitrary choice.\cite{har63} As we show in appendix \ref{altderiv}, this equal sharing of potential energy between particles, regardless of charge, is not accidental.  Without it, energy would be transferred over large distances between otherwise non-interacting (neutral) systems.  

It is worth noting that if $\hat{v}^{eI}_{r_i\tilde{r}_\ell}$ is replaced by a non-local pseudopotential 
\begin{align}
  \hat{v}^{PP}_{r_i\tilde{r}_\ell} = \sum_Y \operator{Y}{v^Y_{r_i\tilde{r}_\ell}}{Y}
\end{align}
in the analysis above (with $\{Y\}$ representing spherical harmonics), the density operator for pseudopotential energy has the same form:
\begin{align}
  \hat{V}^{PP}_r = \sum_{i\ell}\frac{\hat{\delta}_{rr_i}+\hat{\delta}_{r\tilde{r}_\ell}}{2}\hat{v}^{PP}_{r_i\tilde{r}_\ell}.
\end{align}
Thus pseudo-ions are treated on the same footing as other particles.

An alternative form can be introduced which bears some similarity to the classical energy density of electric fields.  If we denote the Coulomb potential due to a point charge $q$ sitting at $r_q$ as $v^q_{rr_q}$ and recall that $-\nabla_r^2v^q_{rr'}=q\delta_{rr'}$, the potential energy of a pair of charges can be written in the following way:
\begin{align}
  v^{qq'}_{r_qr_{q'}} &= \int dr v^{q}_{rr_{q}}q'\delta_{rr_q'} \label{trans1}\\
                   &= -\int dr v^{q}_{rr_{q}}\nabla_r^2v^{q'}_{rr_{q'}} \\
                   &= \int dr \nabla_rv^{q}_{rr_{q}}\nabla_rv^{q'}_{rr_{q'}} - \int ds\cdot v^{q}_{rr_{q}}\nabla_rv^{q'}_{rr_{q'}}. \label{trans3}
\end{align}
If the integral is taken over the entire simulation domain, then the surface term vanishes and we are left with an energy density for the pair:
\begin{align}
  v^{qq'}_r = \nabla_rv^{q}_{rr_{q}}\nabla_rv^{q'}_{rr_{q'}}.
\end{align}
The resulting energy density for a collection of point charges involves products of electric fields from different particles, rather than the square of the total electric field, and thus avoids the infinite energy involved in the formation of a point charge.  Applying the transformations entailed in Eqs. \ref{trans1}-\ref{trans3} to Eq. \ref{totalpot}, a density operator that stores potential energy in these single particle fields is
\begin{align}\label{pdensop_field}
  \hat{V}^{field}_r &= \sum_{i<j}\nabla_r\hat{v}^e_{rr_i}\nabla_r\hat{v}^e_{rr_j} 
                    + \sum_{i\ell}\nabla_r\hat{v}^e_{rr_i}\nabla_r\hat{v}^I_{r\tilde{r}_\ell} \nonumber\\ 
           &\qquad  + \sum_{\ell<m}\nabla_r\hat{v}^I_{r\tilde{r}_\ell}\nabla_r\hat{v}^I_{r\tilde{r}_m}.
\end{align}
It should be noted that this field form assumes a system composed of Coulomb particles.  It is unlikely that similar forms exist for general pair interactions, $e.g.$ dispersive interactions between helium atoms cannot be described by the electric field from a static charge density since they arise from induced correlations between particles.

Besides its potential lack of generality, the field form poses other difficulties for QMC calculations.  First, $\hat{V}^{field}_r$ would have to be evaluated over an entire grid or basis expansion for each sample, whereas $\hat{V}_r$ can be accumulated with a simple histogramming approach.  Second, the divergences in $\nabla_rv^{q}_{rr_{q}}\nabla_rv^{q'}_{rr_{q'}}$ would result in a high variance, since large positive or negative values would be accumulated as any particle passed near a grid point.  Similarly, a basis expansion of $V^{field}_r$ would have difficulty capturing the divergences, potentially introducing bias as well as having a large variance.  For these reasons, we prefer $\hat{V}_r$ for use in QMC.

\subsection{Total Energy Density}
To summarize, the energy density operator we have derived has the following form:
\begin{align}\label{edensop}
  \hat{\mathcal{E}}_r &= \frac{1}{2}\sum_i\hat{\delta}_{rr_i}\left[\hat{p}_i^2+\sum_{j\ne i}\hat{v}^{ee}_{r_ir_j}+\sum_\ell\hat{v}^{eI}_{r_i\tilde{r}_\ell}\right] \nonumber \\
   \qquad &  + \frac{1}{2}\sum_\ell\hat{\delta}_{r\tilde{r}_\ell}\left[\sum_i\hat{v}^{Ie}_{r_i\tilde{r}_\ell}+\sum_{m\ne\ell}\hat{v}^{II}_{\tilde{r}_\ell\tilde{r}_m}\right].
\end{align}
It divides the total energy into single particle contributions, with the particles themselves carrying the kinetic and potential energy (as expressed by the delta functions).

If the energy density operator is integrated over a subvolume of the entire domain, one obtains a Hamiltonian operator which reasonably represents that subsystem.  As an example, consider the Hamiltonian for an infinite number of particles inhabiting all of space.  If the particles are constrained to undergo periodic motion in an array of cells, we have an extended Ewald system with energy density operator
\begin{align}\label{edensinf}
  \hat{\mathcal{E}}^\infty_r &= \frac{1}{2}\sum_c\sum_i\hat{\delta}_{rr_{ci}}\left[\hat{p}_{ci}^2+\sum_{j\ne i}\hat{v}_{r_{ci}r_{cj}}\right]\nonumber\\
  \qquad & + \frac{1}{2}\sum_{c\ne c'}\sum_{ij}\left[\hat{\delta}_{rr_{ci}}\hat{v}_{r_{ci}r_{c'j}}+\hat{\delta}_{rr_{c'j}}\hat{v}_{r_{ci}r_{c'j}}\right],
\end{align}
where $c$ is the cell index, and $r_{ci}$ is the position of the $i$-th particle in cell $c$.
Integrating this energy density operator over a single periodic cell results in the familiar Ewald Hamiltonian involving only the particles in the cell, along with the potential felt from periodic images:
\begin{align}
  \int_{\Omega_0}dr\hat{\mathcal{E}}^\infty_r &= \sum_i \frac{\hat{p}^2_{0i}}{2}+\sum_{i<j}\hat{v}_{r_{0i}r_{0j}}+\frac{1}{2}\sum_{c\ne 0}\sum_{ij}\hat{v}_{r_{0i}r_{cj}}.
\end{align}
In order to calculate the potential energy density resulting from Ewald interactions, special care must be given to separate out single particle energies (including the partitioning of constant terms).  This issue is addressed in more detail in appendix \ref{ewalddens}.

\begin{table}[b]
  \tabcolsep 5pt
  \begin{tabular}{lccc}
    & \mct{Total} & \mct{Kinetic} & \mct{Potential} \\
    \hline\hline \\
   \mct{He isolated  } &  -2.8629(08) &   2.6543(67) &  -5.5172(68) \\
   \mct{He integrated} &  -2.8622(10) &   2.6470(60) &  -5.5093(60) \\
    \hline \\
   \mct{Ne isolated  } & -33.9157(32) &  22.571(25)  & -56.487(25) \\
   \mct{Ne integrated} & -33.9159(26) &  22.566(16)  & -56.482(17) \\
   \hline\hline\\
  \end{tabular}
  \caption{\label{HeNetable}
    Total, Kinetic, and Potential atomic energies for He and Ne from isolated total energy calculation or integrated energy density of He-Ne pair.  All energies are in Hartree units.  Parentheses indicate statistical uncertainty in the last two decimal places.
  }
\end{table}

\section{Demonstration of Separability: A helium-neon ``Gas''}\label{sec_HeNe}

The energy density represents a partitioning of the total energy among particles.  If the energy density provides a physical description of this partitioning, it should match the partitioning naturally found in separable systems.  For example, the atoms in a low density helium-neon gas only weakly interact and so, to a good approximation, the total many body wavefunction separates into a product of atomic wavefunctions, $\Psi_{tot}=\prod_a\Psi_a$.  In this case, the total energy also separates into a sum of atomic energies: 
\begin{align}
E_{tot} = \expvalnh{\Psi_{tot}}{\hat{H}_{tot}}{\Psi_{tot}} \approx \sum_a\expvalnh{\Psi_a}{\hat{H}_a}{\Psi_a} = \sum_aE_a.
\end{align}
Such atomic energies constitute a physically meaningful partitioning of the energy, which should be reflected in the energy density.  Any deviation from this partitioning can be considered a long-ranged transfer of energy between weakly interacting systems.  As shown in appendix \ref{altderiv}, the energy density operator of Eq. \ref{edensop} should be capable of isolating these atomic energies when integrated over single atom volumes.

To test this property, we have performed Variational Monte Carlo\cite{mcmillan1965} simulations of three systems: a single helium atom, a single neon atom, and a helium-neon pair, each using the Ewald summation to represent an infinite gas.  In all cases, the interatomic distance was fixed at $14$ a.u. on a square lattice and pseudopotentials generated with the OPIUM\cite{opium} program represented ion cores (nucleus of helium and the He core of neon).  Orbitals were generated by the Quantum Espresso\cite{gianozzi2009} package with a cutoff of $150$ Ry.  All QMC calculations were performed with QMCPACK.\cite{kim12} The conclusions reached here are insensitive to these choices and generalize to DMC calculations, since DMC can be represented as a reweighting of VMC configurations.

The results of the calculations are shown in Table \ref{HeNetable}.  The integrated energies from the He-Ne pair system match the energies of the single atom systems to within a few mRy, verifying that the energy density operator partitions the energy in a physically meaningful way.

\section{Efficient Energy Differences: $\delta$-Doped silicon}\label{sec_ddSi}

Beyond providing physical insight into local energetics, the energy density is particularly useful in QMC because it can reduce the statistical variance of energy differences.  In some cases the desired quantity can be obtained in a single calculation with the energy density,\cite{che92,yu11} further increasing the efficiency.  To illustrate these points, we have performed a series of DMC energy density calculations for a simple model of $\delta$-doped silicon, where variations in the energy density are expected to vary along only one spatial direction.

In a $\delta$-doped material, a thin layer of dopant atoms is sandwiched between two bulk regions.  Since we are interested only in demonstrating the efficiency enhancement of the energy density over total energy calculations, we introduce several idealizations in our system.  First, the $\delta$ layer is represented by a plane of single atoms.  Second, the ``dopant'' is chosen to be germanium since it is not expected to introduce long-ranged charge disturbances.  Third, the cell dimensions parallel to the $\delta$ layer are chosen minimally; the system is represented as a line of 2-atom primitive cells (8, 12, or 16 cells long) and doping is achieved by replacing a single Si atom with Ge, forming a $(110)$ plane.  These idealizations are made for convenience only and do not affect our conclusions.

\begin{figure}
  \includegraphics[scale=.35]{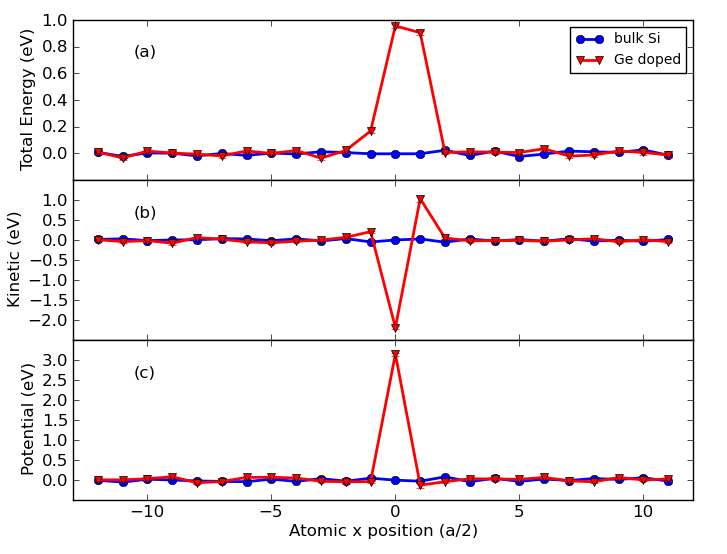}
  \caption{\label{fatomic}Total $(a)$, kinetic $(b)$, and potential $(c)$ atomic DMC energies for $12\times 1 \times 1$ systems. Bulk Si is shown in blue with $\delta$-doped in red.  Atomic positions are projected onto the Cartesian x-axis and shown in units of $a/2$.  Each atom represents an equivalent $(110)$ plane. Ge is at the origin.}
\end{figure}

It is natural to discuss the energy density in terms of atomic energies.  Here volumes assigned to each atom are chosen to be the Voronoi\cite{voronoi1908} polyhedra of the nuclei which enclose the set of points nearest each nucleus.  Though more physically motivated and transferable volumes can be found,\cite{yu11} they depend on the detailed shape of the final density which makes a histogramming approach difficult.  Also, in many cases the region of interest is joined to a bulk region where any repeated volumes of the appropriate size will contain the same energy due to the periodicity of the lattice.  Thus differing definitions in atomic volumes can only change the location of the boundary enclosing the bulk by no more than a single primitive cell.

Total, kinetic, and potential atomic energies for the 24-atom system ($12\times1\times1$ tiled primitive cell at the gamma point) are shown in Fig. \ref{fatomic}.  The energy profiles are asymmetric due to the bipartite diamond lattice.  As expected, atomic energies for the $\delta$-doped system relax to bulk values away from the dopant plane, further establishing the local quality of the energy density operator (see Eq. \ref{edensop}).

In Diffusion Monte Carlo, the total energy is estimated more accurately than other quantities, since the mixed estimator provides the exact answer, $E^{mix}=\expval{\Psi_0}{H}{\Psi_T}/\overlap{\Psi_0}{\Psi_T}=E_0$.  Observables ($\mathcal{O}$) that do not commute with the Hamiltonian are estimated using the extrapolated estimate, $\mathcal{O}^{extrap}\equiv 2\mathcal{O}^{mix}-\mathcal{O}^{trial}$, which is accurate to second order in the trial wavefunction error.  Here $\mathcal{O}^{trial}$ is obtained from a separate VMC calculation.  

The energy density presents an interesting situation: although it is intimately related to the total energy, the energy density operator does not commute with the Hamiltonian.  For very small volumes ($\delta\Omega$), it is reasonable to expect the mixed estimate to be more accurate, since $\int_{\delta\Omega}dr\hat{\mathcal{E}}_r$ and $\hat{H}$ will be most different.  It is also clear that for volumes ($\Omega'$) approaching the system size, the commutator of $\int_{\Omega'}dr\hat{\mathcal{E}}_r$ and $\hat{H}$ tends to zero.  Thus there will always be some integration volume for which the mixed estimator is more accurate.

For cases similar to our $\delta$-doped Si example, the right choice for calculating total energy differences from energy densities is the mixed estimator.  Since the DMC total energy is correct for the bulk system, the bulk atomic energies are also correct due to symmetry.  The same conclusion is reached for atomic energies in the bulk-like regions of the $\delta$-doped system, since they match the energies of the bulk system (see Fig. \ref{fatomic}).  Thus the mixed estimator is also exact for the dopant region and the extrapolated value underestimates the exact value by the VMC error.

\begin{figure}
  \includegraphics[scale=.35]{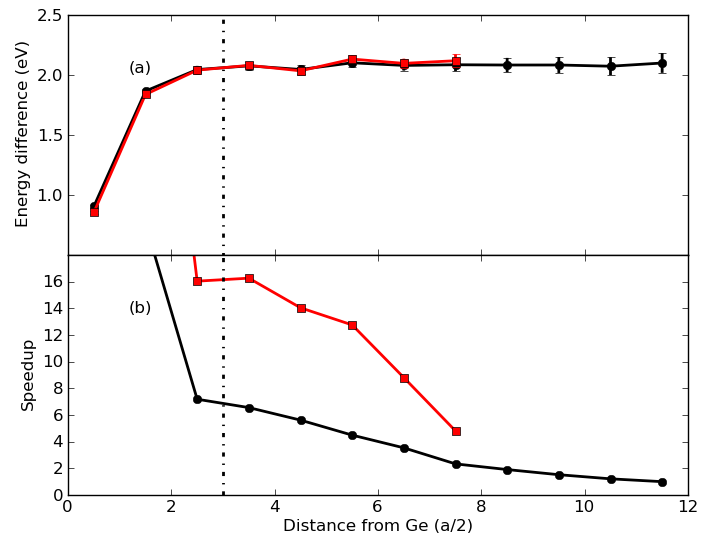}
  \caption{\label{fint}Energy difference $(a)$ and speedup $(b)$ vs. integration distance from Ge dopant.  Energy differences between the doped and bulk systems are marked in black.  Quantities in red use the bulk-like region of the doped system to estimate bulk energies.  Speedup is relative to standard total energy differences.}
\end{figure}

Due to the cancellation of atomic energies in the bulk-like regions, the energy difference of interest can be written as
\begin{align}
  E^{doped}_{tot}-E^{bulk}_{tot} &= \int_{\Omega_d}dr\mathcal{E}_r^{doped}-N_{d}E^{bulk}_a,
\end{align}
where the volume $\Omega_d$ encloses all energy disturbances due to the $\delta$ layer, $N_d$ is the number of atoms within $\Omega_d$, and $E^{bulk}_a$ is the energy per atom of bulk.  As in the standard approach, $E^{bulk}_a$ can be computed in a separate bulk calculation, but the energy density provides an alternative.  It can also be estimated from the bulk-like region of the doped system,
\begin{align}
  E^{bulk}_a &= \frac{1}{N_b}\int_{\Omega_b}dr\mathcal{E}_r^{doped},
\end{align}
with $\Omega_b$ and $N_b$ satisfying $\Omega_d\cup\Omega_b=\Omega_{tot}$ and $N_d+N_b=N_{tot}$, respectively.  Thus the energy difference can be obtained from a single calculation.

Figure \ref{fint}$a$ shows the convergence of the energy difference as the integration volume surrounding the $\delta$ layer is increased.  The final point in black corresponds to the standard total energy difference.  Consistent with Fig. \ref{fatomic}, the energy difference converges to its final value once the neighboring layers are summed over.  

\begin{table}[b]
  \tabcolsep 5pt
  \begin{tabular}{lccc}
    & $8\times 1\times 1$ & $12\times 1\times 1$ & $16\times 1\times 1$\\
    \hline\hline \\
   \mct{doped \& bulk } &  3.5  &   6.6  &   8.9  \\
   \mct{doped only}     &  9.3  &  16.3  &  22.0 \\
   \hline\hline\\
  \end{tabular}
  \caption{\label{tsize}
    Speedup at fixed integration volume for various system sizes ($M\times 1 \times 1$ primitive cell tilings).  
  }
\end{table}

The statistical variance is also greatly reduced relative to total energy differences, resulting in a significant speedup as shown in Fig. \ref{fint}$b$.  The speedup is defined as the ratio computer times required to reach the same statistical error using the two approaches.  For the smallest converged volumes, the computational cost is diminished by a factor of 16.  In general the speedup achieved depends on the relative size of the subsystem of interest.  For a system of nearly uniform composition with $N_{tot}$ atoms and a subsystem of $N_{sub}$ atoms, it will roughly scale as $N_{tot}/N_{sub}$.

\begin{figure}
  \includegraphics[scale=.35]{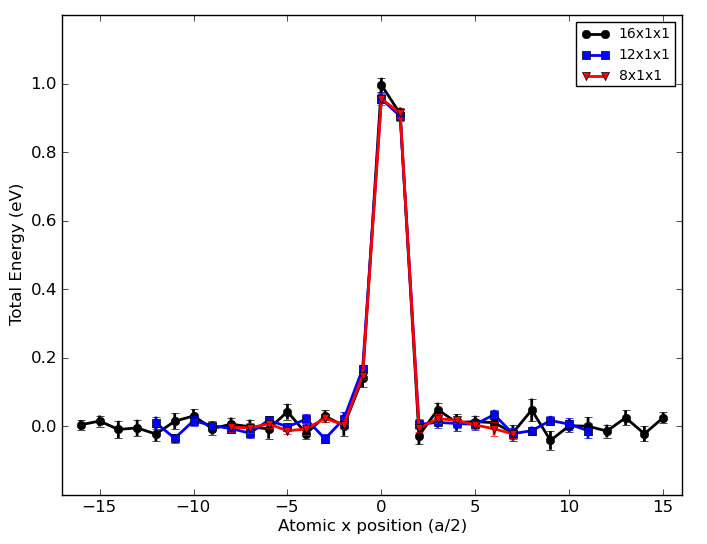}
  \caption{\label{fsize}Atomic DMC energies for 16, 24, and 32 atom $\delta$-doped Si vs atomic position.  Ge is at the origin.}
\end{figure}

This can amount to a substantial savings, since a series of calculations of increasing size are often performed to extrapolate to the thermodynamic limit.  To represent this process, Fig. \ref{fsize} displays atomic energies for the 16, 24, and 32 atom $\delta$-doped systems.  For our model system, finite size effects due to confinement and artificial periodicity are rather small (along the $[110]$ direction only), as evidenced by the similarity of the dopant energy profiles.  The same integration volume can therefore be used throughout to obtain the energy differences.  The speedup gained by the energy density approach in each case is summarized in Table \ref{tsize}.

\section{Comparison to DFT Energy Density}\label{sec_dft}
Though Density Functional Theory (DFT) originated with the mapping of the ground state density onto the total energy, its use has been extended to include a variety of other physical observables.  Since many these properties represent an approximation beyond the use of inexact functionals (\emph{i.e.}, their accuracy depends on the overall physical relevance of the Kohn-Sham system itself), their fidelity to the true system must be assessed via benchmark calculations.  The addition of the energy density to this repertoire should be given the same comparison.
 
\begin{figure}
  \includegraphics[scale=.45]{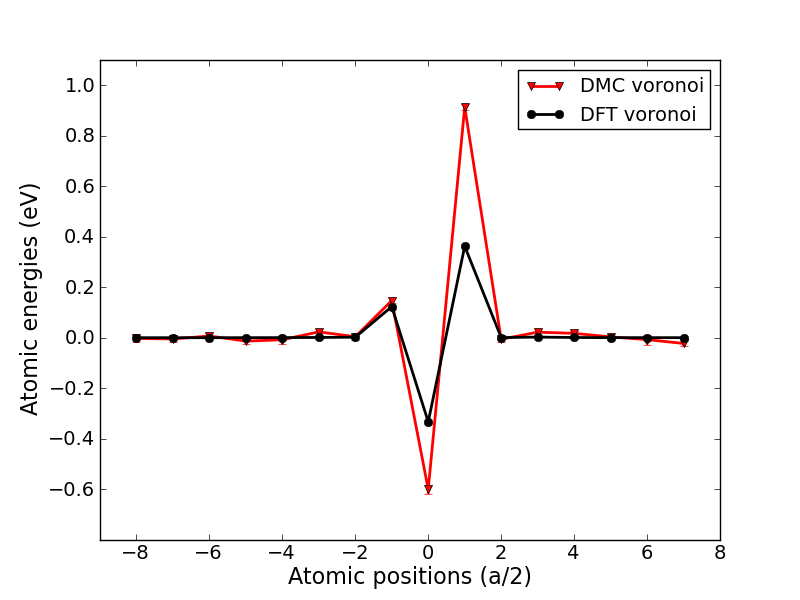}
  \caption{\label{fdft}Atomic energies referenced to respective bulk values of the $8\times 1\times 1$ doped system. Results for DMC are shown in red and DFT in black.  Ge is at the origin.}
\end{figure}

Since the DFT total energy must derive from the many-body Hamiltonian, the transformations found earlier can be used to obtain a DFT energy density.  We will start with the expression for the DFT total energy as defined in Yu, \emph{et al.}:\cite{yu11}
\begin{align}\label{DFT_total_energy}
  E_{tot}^{DFT} &= T+V_{ee}+V_{eI}+V_{II}\\
  T &= -\frac{1}{2}\sum_{n}\int dr \phi_{nr}^*\nabla_r^2\phi_{nr} \\
  V_{ee} &= \frac{1}{2}\int dr dr' \rho_r v^C_{rr'}\rho_{r'} + \int dr \rho_r \epsilon_{XC}(\rho_r,\nabla_r\rho_r) \\
  V_{eI}  &= \int dr\sum_\ell\left[\rho_rV^L_{rr_\ell} + \sum_{nY}\phi_{nr}^*V^{NL}_{Yrr_\ell}\hat{P}_Y\phi_{nr}\right] \\
  V_{II}  &= \sum_{\ell<m}Z_\ell Z_mv^C_{\ell m},
\end{align}
where $r$ \& $r_\ell$ are the electronic and ionic coordinates, $\phi_{nkr}$ and $\rho_r$ are the DFT orbitals and density, $v^C$ is the bare Coulomb energy, and $V^L_{rr_\ell}$ and $V^{NL}_{Y,rr_\ell}$ are the local and non-local parts of the pseudopotential centered on ion $\ell$.

Applying the transformations
\begin{align}
  \nabla^2_{r_1}&\rightarrow\delta_{rr_1}\nabla^2_{r_1}\\
  v^C_{r_1r_2}&\rightarrow\frac{\delta_{rr_1}+\delta_{rr_2}}{2}v^C_{r_1r_2}\\
  V^{L/NL}_{r_1r_\ell}&\rightarrow\frac{\delta_{rr_1}+\delta_{rr_\ell}}{2}V^{L/NL}_{r_1r_\ell}
\end{align}
to Eq. \ref{DFT_total_energy}, we find expressions for DFT kinetic and potential energy densities associated with either the electrons or ions:
\begin{align}
  T^e_r &= -\frac{1}{2}\sum_{n}\phi^*_{nr}\nabla^2_r\phi_{nr}\\
  V^e_r &= \rho_r\left[ \frac{1}{2}\int dr' v^C_{rr'}\rho_{r'} + \epsilon_{XC}(\rho_r,\nabla_r\rho_r) + \frac{1}{2}\sum_\ell V^L_{rr_\ell} \right] \nonumber\\
        &\qquad + \frac{1}{2}\sum_{\ell nY}\phi_{nr}^*V^{NL}_{Yrr_\ell}\hat{P}_Y\phi_{nr}\\
  V^I_r &= \sum_\ell \delta_{rr_\ell}\Bigg[ \frac{1}{2}\sum_{m\ne \ell}Z_\ell Z_mv^C_{\ell m}  + \frac{1}{2}\int dr'\rho_{r'}V^L_{r'r_\ell} \nonumber \\
    & \qquad + \frac{1}{2}\sum_{nY}\int dr'\phi_{nr}^*V^{NL}_{Yr'r_\ell}\hat{P}_Y\phi_{nr} \Bigg].
\end{align} 

For comparison, the energy density of Yu, \emph{et al.}\cite{yu11} is
\begin{align}\label{Martin_form}
  T^e_r &= -\frac{1}{2}\sum_{n}\phi^*_{nr}\nabla^2_r\phi_{nr}\\
  V^e_r &= \rho_r\left[ \frac{1}{2}\int dr' v^C_{rr'}\rho_{r'} + \epsilon_{XC}(\rho_r,\nabla_r\rho_r) + \frac{1}{2}\sum_\ell V^L_{rr_\ell} \right] \\
  V^I_r &= \sum_\ell \rho^L_{rr_\ell}\left[ \frac{1}{2}\sum_{m\ne \ell}V^L_{rr_\ell} + \frac{1}{2}\int dr'\rho_{r'}V^L_{r'r_\ell}\right] \nonumber\\
    &\qquad + \sum_{\ell}\delta_{rr_\ell}\sum_{nY}\int dr'\phi_{nr}^*V^{NL}_{Yr'r_\ell}\hat{P}_Y\phi_{nr}, 
\end{align} 
where $\rho^L_{rr_\ell}$ is the classical charge density corresponding to the local part of the pseudopotential, $\rho^L_{rr_\ell}\equiv -\nabla^2V^L_{rr_\ell}$.  Here the model potential and density employed by Yu \emph{et al.}\cite{yu11} have been removed, as they were introduced for numerical accuracy only.    

The only distinctions between the two forms, as represented above, involve the treatment of pseudopotentials.  First, the form of Yu \emph{et al.}\cite{yu11} locates all of the energy from the non-local part of the pseudopotentials on the ion cores rather than distributing half of it to the electrons within the cutoff radius.  Second, the portion of the electron-ion energy belonging to ion $\ell$ is distributed according to $\rho^L_{rr_\ell}$ rather than as a delta function, though this charge density approaches zero as the local potential approaches $-Z_\ell/\abs{r-r_\ell}$.  Provided the region of interest contains one or more atomic volumes, either definition of the DFT energy density will yeild quantitatively similar results.  Thus results obtained from the VASP\cite{kresse1993,kresse1994,kresse1996a,kresse1996b} implementation of Eqs. \ref{Martin_form}, generalized to the PAW formalism,\cite{blochl1994,kresse1999,yu11} can be meaningfully compared to our DMC calculations.

The DFT (PBE\cite{perdew1996,perdew1997}) energy density of the doped system is shown in Fig. \ref{fdft} alongside our DMC results.  Despite the proximity of neighboring particle images in the transverse direction, test DMC calculations in a larger 108 atom system showed that finite size effects due to correlation are small.  The model potential and density in the formalism of Yu \emph{et al.}\cite{yu11} introduces an independent reference energy for each atomic species, and so bulk values of Si or Ge have been subtracted from each respective atomic energy in the doped system.  Both PBE and DMC agree regarding the range of the energy disturbance due to Ge and whether atomic energies lie above or below the corresponding bulk values.   However the magnitude of the departure from bulk differs by as much as a factor of two.  A remaining source of ambiguity in comparing DMC and DFT energy densities is the non-uniqueness of the terms involving exchange and correlation: $-\frac{1}{2}\sum_{n}\int dr \phi_{nr}^*\nabla_r^2\phi_{nr} +\rho_r\epsilon_{XC}(\rho_r,\nabla_r\rho_r)$.  Although functionals that reproduce the exact energy density will also produce the exact ground state energy by default, this requirement represents an additional constraint beyond the canonical density and total energy.

\section{Conclusion}\label{sec_conclusion}
Consistent with prior studies, we have derived an energy density operator that provides a description of local energetics.  This operator is well suited for QMC calculations as it can be implemented in existing codes with minimal effort and without adding to the computational cost.  For the case of Diffusion Monte Carlo, a simple asymmetric form of the kinetic energy density operator can be used as long as the wavefunction is real, or, if twist averaging, the k-point set has inversion symmetry.  We have also established that standard DMC mixed estimates of subsystem energies should be preferred over extrapolated estimates when calculating total energy differences, despite the non-commutativity between total and subsystem energy operators. 

We have demonstrated that the energy density can be used to improve the efficiency of energy difference calculations.  The speedup realized in a given calculation is related to the degree of cancellation between total energies.  Though the gains will depend on the physical system, we anticipate that the energy density approach will be worthwhile for many important systems, such as surfaces, interfaces, and point defects.

\section*{Acknowledgements}
The authors are indebted to Richard Martin and Dallas Trinkle for many thoughtful discussions providing useful perspective on the quantum energy density.  This work was supported by the National Science Foundation (NSF) under No. 0904572 and by the  DOE-BES Materials Sciences and Engineering Division and the Laboratory Directed Research and Development Program of Oak Ridge National Laboratory, managed by UT-Battelle, LLC. Computer time was provided by the DOE-INCITE and NSF-Teragrid programs.

\appendix

\section{Alternative Derivation of the Energy Density Operator}\label{altderiv}
The energy density operator $\hat{\mathcal{E}}_r$ can be derived by starting with a general form and imposing successive physical conditions.  Specifically, $\hat{\mathcal{E}}_r$ must integrate to the Hamiltonian $(\int dr\hat{\mathcal{E}}_r=\hat{H})$ and will be constrained to share its symmetries.  It will also be required to obey an isolation principle: neutral atoms far away from each other do not exchange energy.  As the energy density operator is not unique, one condition will involve a conscious choice of representation and hence is a matter of taste or convenience.

The real space form of the energy density operator can be written as
\begin{align}\label{edgen}
  \hat{\mathcal{E}}_r = \sum_ia^i(r,R)\hat{h}_i+\sum_{i<j}b^{ij}(r,R)\hat{v}_{ij},
\end{align}
where $\hat{h}_i$ is the single particle Hamiltonian of particle $i$ and $\hat{v}_{ij}$ is the potential between particles $i$ and $j$.  The functions $a^i_{rR}$ and $b^{ij}_{rR}$ distribute the energy from each term over space $(r)$, possibly depending on all particle coordinates $(R)$.  In all expressions, $i$ labels a particle, while the composite index $ij$ labels a bond, and so $ij$ and $ji$ are interchangeable.  

The non-commutativity of $\hat{h}_i$ and $a^{i}_{rR}$ is ignored, since a more general representation only leads to an arbitrary linear combination of $\hat{T}^{L^2+R^2}_r$,  $\hat{T}^{LR}_r$, $\hat{T}^{mod}_r$, or other equivalent forms adding unnecessary clutter to what follows.  As before, we select $\hat{T}^{R^2}_r$ for practical reasons, which Eq. \ref{edgen} reflects.

The terms already present in a Hamiltonian do not change form when a new particle is added to the system.  In particular, they do not depend on the position or momentum of the new particle.  The final form of the Hamiltonian also does not depend on the order in which particles are added.  Imposing these invariances on the energy density constrains the distribution functions to depend only on the particle coordinates of the corresponding Hamiltonian terms:
\begin{align}
  a^i(r,R) = a^i(r,r_i) \qquad b^{ij}(r,R)=b^{ij}(r,r_i,r_j).
\end{align}

Next, we specify where the energy is to be stored.  The energy can be stored in fields of various kinds, such as $b^{ij}_{rr_ir_j}=\nabla_rv_{rr_i}\nabla_rv_{rr_j}/v_{r_ir_j}$, as we have seen before.  However, we will view the energy as being carried by the particles:
\begin{align}
  a^i(r,R) &= a^i(r_i)\delta_{rr_i} \\
  b^{ij}(r,R) &= b^{ij}_i(r_i,r_j)\delta_{rr_i}+b^{ij}_j(r_i,r_j)\delta_{rr_j}.
\end{align}

The physical system is invariant under the translation or inversion of all particles ($r_i\rightarrow r_i+\Delta r$ and $r_i\rightarrow -r_i$), so
\begin{align}
  a^i(r,R) &= a^i\delta_{rr_i} \\
  b^{ij}(r,R)&=b^{ij}_i(r_i-r_j)\delta_{rr_i}+b^{ij}_j(r_i-r_j)\delta_{rr_j}\\
  b^{ij}_{i/j}(-r) &= b^{ij}_{i/j}(r). 
\end{align}

Since the energy density merely redistributes energy, we have the normalization condition $\int dr\hat{\mathcal{E}}_r=\hat{H}$, which further requires
\begin{align}
  a^i=1 \qquad b^{ij}_i(r_{ij})+b^{ij}_j(r_{ij})=1.
\end{align}

The energy density operator must also be invariant under the exchange of identical particles ($r_i\leftrightarrow r_j$):
\begin{align}
  b^{ij}_i(r_{ij}) = b^{ij}_j(r_{ij}) = \frac{1}{2} \qquad \textrm{(identical particles)}.
\end{align}

To establish the same partitioning of interaction energy among disparate particles, consider a system comprised of a nucleus $n$ of charge $Z$ and $Z+1$ electrons.  Integrating the pair potential energy density operator over a volume containing only electron $Z+1$,
\begin{align}
  \hat{v}_{Z+1} &= \int_{\Omega_{Z+1}} dr \hat{V}^{pair}_r \\
               &= \frac{1}{2}\sum_{i=1}^Z\frac{1}{\abs{r_{Z+1}-r_i}} - \frac{Zb^{en}_e(r_{Z+1}-r_n)}{\abs{r_{Z+1}-r_n}}.
\end{align}
Since the unknown function $b^{en}_e$ depends only on the positions of the nucleus and electron $Z+1$, we are free to consider any configuration of the first $Z$ electrons.  By moving them arbitrarily close to the nucleus, the interaction energy between the electron and the compressed neutral atom must vanish:
\begin{align}
  \hat{v}_{Z+1}\rightarrow \frac{Z}{\abs{r_{Z+1}-r_n}}\left(\frac{1}{2}-b^{en}_e(r_{Z+1}-r_n)\right) = 0.
\end{align}
This can only be satisfied if $b^{en}_e(r_{Z+1}-r_n)=1/2$ and thus $b^{en}_n(r_{Z+1}-r_n)=1/2$ also.  

Though the configuration used to establish this point is highly improbable in nature, any deviation from an equal partitioning of energy has real physical consequences.  If the electron and atom were a large distance apart (far enough so that dispersive forces are negligible, but Coulomb forces are not), the same conclusion is reached since the atomic size is small compared to the separation distance.  The equal partitioning guarantees that the energy shared with the distant electron by the electron cloud and nucleus exactly cancel.  If this were not the case, the energy attributed to a neutral atom far from a neutral chunk of matter would vary like $1/r$, causing a long ranged energy transfer between effectively isolated systems. 

Since any pairing of electrons and/or protons ($e-e$, $e-p$, or $p-p$) results in each particle carrying $1/2$ of the interaction energy, it immediately follows that a pair of composite particles (such as nuclei with differing charge) also evenly split interaction energy.  This line of reasoning also extends to coarse grained models of atomic interactions, such as helium atoms interacting via an approximate pair potential.  Thus if particles carry the energy, the energy density operator must have the form
\begin{align}
  \hat{\mathcal{E}}_r = \sum_i\frac{\delta_{rr_i}\hat{h}_i+\hat{h}_i\delta_{rr_i}}{2}+\sum_{i<j}\frac{\delta_{rr_i}+\delta_{rr_j}}{2}\hat{v}_{ij},
\end{align}
which has been symmetrized to restore hermiticity.

\section{The Ewald Potential Energy Density}\label{ewalddens}
To compute the energy density according to Eq. \ref{edensop}, the energy carried by a single particle must be determined.  In the standard breakup of the Ewald potential\cite{ewald1921,natoli1995} into long and short ranged parts, it is not immediately obvious how constant terms in the potential energy should be distributed among particles.  Beginning with the general description of energy density in an infinite Ewald system (Eq. \ref{edensinf}), we determine the correct partitioning.

The single particle energy is isolated by integrating the energy density around particle $i$ in an arbitrary cell:
\begin{align}
  v_i^{Ewald} &= \frac{1}{2}\sum_{j\ne i}v_{ij}(r_{ij})+\frac{1}{2}\sum_{L\ne 0}\sum_jv_{ij}(r_{ij}+L).
\end{align} 
The set of vectors $\{L\}$ mark the centers of the infinite cell array. 
Dividing each Ewald pair potential into long and short ranged components,
\begin{align}
  \sum_L v_{ij}(r+L) = v^s_{ij}(r)+v^\ell_{ij}(r),
\end{align}
and denoting corresponding Fourier transforms as $\tilde{v}^s_{ij}(k)$ and $\tilde{v}^\ell_{ij}(k)$, we find the desired single particle energy:
\begin{align}
  v_i^{Ewald} =\frac{1}{2}\sum_{j\ne i}&\Big[v^s_{ij}(r_{ij})-\tilde{v}^s_{ij0}+\sum_{0\le\abs{k}\le k_c}e^{ikr_{ij}}\tilde{v}^\ell_{ijk}\Big] \nonumber\\
  +\frac{1}{2}&\Big[-\tilde{v}^s_{ii0}-v^\ell_{ii}(0)+\sum_{0\le\abs{k}\le k_c}\tilde{v}^\ell_{iik}\Big].
\end{align}
The standard expression for the total Ewald potential is recovered by summing over all particles.  Similarly, the Ewald potential energy density is the sum of single particle densities:
\begin{align}
  \hat{V}^{Ewald}_r = \sum_i \delta_{rr_i}v_i^{Ewald}.
\end{align}

\bibliographystyle{h-physrev}
\bibliography{ed}

\end{document}